\documentclass{an}
\usepackage{graphicx}
\usepackage{times}
\usepackage{fancyhdr}
\begin{document}

\title{QPO as the Rosetta Stone for understanding black hole accretion}

\author{Marek A. Abramowicz$^{1,2,3}$ et al.}
\institute{
        Corresponding Fellow, Nordita, Copenhagen, Denmark,
	\and
	Visitor at SLAC, Stanford University, and at KITP, UCSB, U.S.A.,
	\and
        Permanent at the Department of Physics, G\"oteborg University            
	S 412 96 G\"oteborg, Sweden}

\date{Received; accepted; published online}

\abstract{
Quasi periodic oscillations (QPO) seen in the X-ray fluxes of individual neutron stars and black hole sources are one of most intriguing phenomena in today's astrophysics. The QPO nature is visibly determined by super-strong Einstein's gravity. I argue here that it also profoundly depends on the MRI turbulence in accretion flows. Understanding the QPO physics may therefore guide accretion theory out of its present state of confusion. 
}
\correspondence{marek@fy.chalmers.se}
\maketitle
\section{Accretion theory in search of a paradigm}  
\label{sec01}
Accretion provides energy for radiation in many astrophysical sources: 
stellar binaries, active galactic nuclei, and proto-planetary disks. 
The black hole accretion in quasars is most powerful and most efficient 
steady energy source known in the Universe. In low-mass X-ray binaries 
(LMXRBs) the compact object, i.e. a neutron star (NS) or a black hole 
(BH) accretes mass lost by its binary companion, i.e. a normal, low 
mass star, $M \sim M_{\odot}$. BH sources in LMXRB are in many 
respects a scaled-down version of quasars, and they are often 
called microquasars\footnote{See an excellent non-technical review by Lasota(1999)}. 

Accretion theory experiences today a period of confusion: a picture of violently unsteady accretion on all its length and time scales emerging from su\-per\-computer simulations, is plainly uncompatible with the long-lasting, quasi-steady accretion states (Shakura-Sunyaev, slim, adaf) predicted by semi-analytic methods. 
\subsection{Shakura-Sunyaev disks, slim disks and adafs}
According to the semi-analytic un\-der\-stan\-ding, developed over the past thirty years, the initially high angular momentum (of matter lost by the low mass star in an LMXRB) is gradually removed by viscous stresses that operate against the shear in the flow. This allows matter to gradually spiral down into the compact object, with gravitational energy degraded to heat also by viscous stresses. A part of the heat converts into radiation, escapes, and cools down the flow. 

When radiative cooling is efficient, a {\it Shakura-Sunyaev}\footnote{The classic work by Shakura \& Sunyaev (1974) is one of the most often quoted papers in modern astrophysics.} type of accretion flow is formed. It is optically thick but geometrically thin (disk-like) and made of relatively cold gas (with negligible radiation pressure) that goes down on very tight spirals, resembling almost circular, almost free (Keplerian) orbits. Shakura-Sunyaev thin disks are relatively luminous, and have ``thermal'' electromagnetic spectra, i.e. not much different from that of a black body.

Radiatively inefficient flows (RIF) are of two types --- the first being a {\it slim disk}\footnote{Abramowicz, Czerny, Lasota \& Szuszkiewicz (1988)}, which is geometrically ``slim'' i.e. rather thin than thick, optically thick, and radiation pressure supported --- and the second being an {\it adaf}\,\footnote{Abramowicz \& al.; (1995); Narayan \& Yi (1995).} which is very hot, optically thin, and geometrically extended, similar in shape to sphere (or ``coronae'') rather than disk. Adaf emits a ``power-law'' non-thermal radiation, often with a strong Compton component. Slim disks are consistent with accretion rates of the order and higher than the Eddington rate, and Shakura-Sunyaev, and adaf are always sub-Eddington. 
\begin{figure*}
\resizebox{\hsize}{!} 
{\includegraphics[angle=0, width=\hsize]{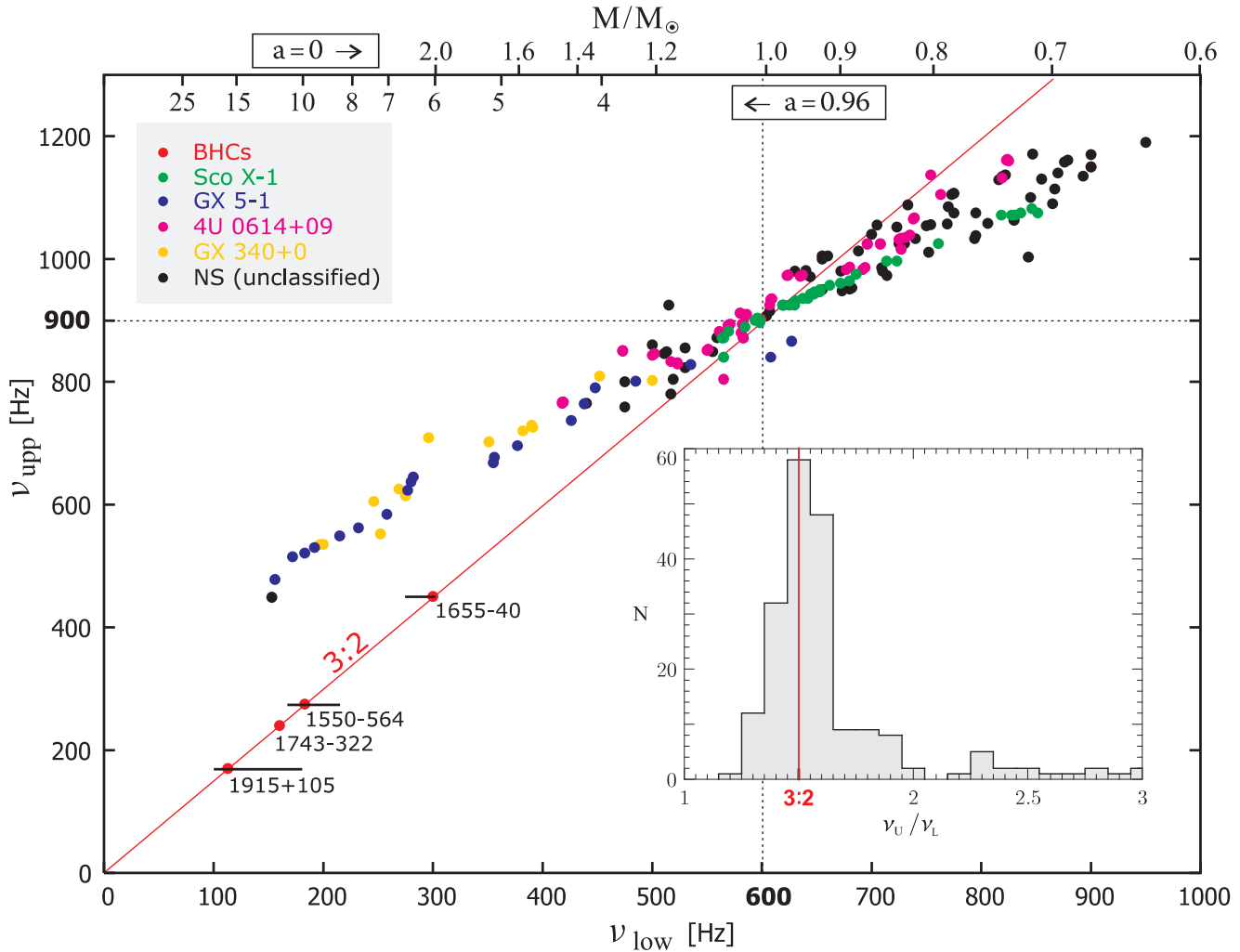}}
\caption{ 
The above ``Bursa plot'' illustrates a rather encouraging qualitative agreement between the twin peak QPO observations and the theoretical model based on a non-linear, strong-gravity resonance. Characteristically for a non-linear resonance the observed ratios of frequencies change in time but stay near the $3/2$ value, determined directly by the strong gravity. For NS sources the frequencies locate approximately along the line $\nu_{\rm up} = a_0\,\nu_{\rm lo} + b_0$. This behavior is generic, and predicted by exact  mathematics of the model. The strong-gravity $1/M$ scaling is also clearly apparent --- see three black horizontal error bars that show measured masses of three microquasars (related to the upper horizontal mass axis). The NS QPO frequencies are about ten times higher than the BH frequencies, because NS masses are smaller by about the same factor.}
\label{fig01} 
\end{figure*}
\noindent Theory predicts that Shakura-Sunyaev, slim disks and adafs are stable: their characteristic steady states are much longer than the dynamical (Keplerian) times and comparable with thermal or viscous times at which a limit-cycle behavior may occur. For the same (sub-Eddington) accretion rate, Shakura-Sunyaev and adaf may simultaneously coexist at different radii or even overlap at some, and the flow may transient between these two types due to some external instabilities (e.g. evaporation). In addition to the accretion inflow, outflows are present, in microquasars often in the form of transient, well collimated jets that shot out with speeds approaching that of light. Jets acceleration and collimation are, most probably, governed by magnetic fields, but relevant details are unknown. It may be that the microquasars jets are powered by the Blandford-Znajek (1977) mechanism, an electromagnetic version of the Penrose process that utilizes BH's rotational energy (see also Li \& Paczy{\'n}ski, 2003).
\subsection{The MRI instability}
In a weakly magnetized fluid, with magnetic field $B_z$ and the Alvf{\'e}n speed $v_A^2 = B_z^2/(4\pi\rho)$ the dispersion relation for perturbations of the fluid quantities $\delta X \sim \exp [{\it i}(kz -\omega t)]$ reads
$\omega^4 - (2kv_A + \omega_r^2)\omega^2 + kv_A (kv_A + r d\Omega^2/dr)=0$,
where $\Omega$ is the angular velocity in the fluid, and $\omega_r$ is the radial epicyclic frequency. Obviously, this equation has an unstable solution $\omega^2 < 0$, if and only if, $kv_A + r d\Omega^2/dr<0$. This is the celebrated magneto rotational instability (MRI), introduced to physics of accretion flows by Balbus \& Hawley (1991; review 1998). MRI generates turbulence. Turbulent stresses dissipate orbital energy and redistribute angular momentum, thus allowing accretion to occur --- without MRI there would be no accretion! However, in my opinion, this fundamental progress in understanding ``from first principles'' the nature of stresses in accretion flows is unmatched by present-day supercomputer magnetohydrodynamical (MHD) simulations. In these simulations, accretion flows appear to violently vary at all time and length scales, never approaching a steady state. 
\subsection{The need for a new paradigm}
The picture of violent accretion is obviously in a direct conflict with the Shakura-Sunyaev, slim disk and adaf models that predict long lasting and orderly quasi-steady states. There is no consensus on what does this acute rift between semi-analytic models and numerical simulations imply for our understanding of accretion. 

I agree with these who are convinced that the picture of quasi-steady accretion must be generally correct, and who point out several shortcomings of numerical simulations today. Simulations still depend too much on technical details and methods used (types of numerical boundary conditions, grid size), and results obtained by different groups do not agree in several relevant details. The simulations do not include radiative processes, and make no predictions about spectral states of accretion flows. Despite their potential and trully impressive sophistication, they have not yet matured, and a lot of work is needed before they will. 

These who believe in MHD simulations, point out that while semi-analytic models will always depend on ad hoc phenomenological parameters to describe stresses, the MHD simulations calculate stresses directly from well understood piece of fundamental physics of MRI.

All in all, understanding of accretion is today somehow confused. Theory makes no sufficiently precise predictions to allow sharp observational judgements, and observations do not combine into a clear qualitative scheme. Phenomenology of spectral states that are really observed in LMXRBs is painfully complex. Neither it agrees with the ``analytic paradigm'' based on the three standard components (Shakura-Sunyaev---adaf or corona---jet), nor there is any indication that it could be explained in foreseeable future\footnote{J.-P. Lasota's estimate: ``Maybe in ten years''} by the ``numerical MRI paradigm''. 

Major puzzles here are: what are the spectral states, what causes transitions between states, how transitions across the ``jet line'' help to launch a jet, and what is the physical reason for the connection between QPO and spectral states. 

In my opinion the last question is the key one. I am convinced that an answer to it would became the Rosetta Stone for accretion theory.
\begin{figure}
\resizebox{\hsize}{!} 
{\includegraphics[angle=-90]{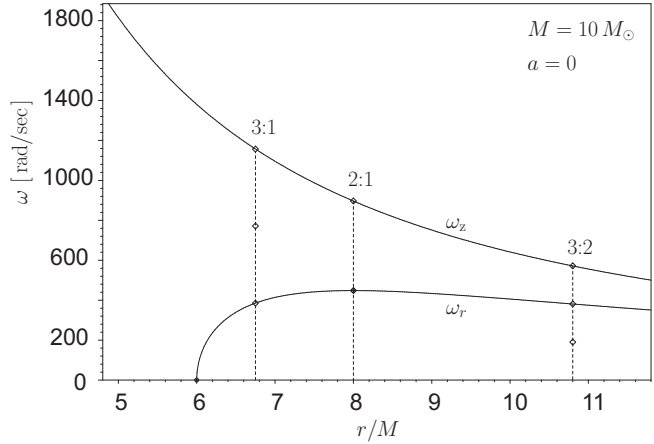}}
\caption{ 
The epicyclic frequencies in Schwarzschild spacetime}
\label{fig02} 
\end{figure}
\section{The {\bf 3/2} twin peak QPO in LMXRBs} 
\label{sec02}
Several of the observational facts discussed in this Section are illustrated in Figure \ref{fig01}, designed by Bursa (2003, unpublished). The Fourier power density spectra (PDS) of the observed LMXRBs X-ray variability reveal a broad band $1/f$ noise, probably consistent with the MRI turbulence, and a few clear QPO peaks with quality factor $Q > 10$. Advanced data analysis points even to $Q > 100$. The highest observed QPO frequencies are in the kHz range, corresponding to Keplerian orbital frequencies a few gravitational radii away from the central NS or BH. In many cases, ``twin peak QPO'' are present with {\it two} characteristic frequencies, $\nu_{\rm up}$ and $\nu_{\rm lo}$, that have frequency ratio close to the $\nu_{\rm up}/\nu_{\rm lo} = 3/2$ value\footnote{The $3/2$ ratio for BH was first noticed, and its importance first stressed, by Abramowicz \& Klu{\'z}niak (2001), and for NS by Abramowicz \& al. (2003). Rational ratios were anticipated by Klu{\'z}niak \& Abramowicz (2000).}. The BH twin peak QPO have frequencies fairly fixed, with ratios being sharply at $3/2$. The frequencies scale inversely with the BH mass\footnote{The $1/M$ scaling was first found by McClintock \& Remillard (2003) for the three microquasars with known masses (see Fig. \ref{fig01}).}. The NS twin peak QPO frequencies wander in time in a wide range, but their changing ratios cluster remarkably sharply at $3/2$, the same value as in the BH case. The NS data shows also the same $3/2$ ratio in the anticorrelation (predicted by the resonance model) between slopes and shifts of the Bursa lines for individual NS sources.
\section{Epicyclic resonance model} 
\label{sec03}
\subsection{Observational motivations and simple physics}
The twin peak BH QPO are very accurate, high frequency, two coupled transient clocks, with frequencies fairly fixed for a particular source from one QPO event to another, that in addition scale from source to source as $1/M$. In my opinion, any serious theoretician is forced to conclude from these facts that the QPO clock is determined by Keplerian motion in strong gravity, and {\it not} e.g. by MRI or some radiative processes. The simplest and the most natural possibility is that the clock mechanism is governed by small {\it epicyclic} deviations $\delta r$, $\delta z$ from the exact circular ($r=r_0$) and planar ($z=0$) motion, as indeed expected in the Shakura-Sunyaev disk. In the zeroth-order approximation, the model thus explains QPO in terms of two simple harmonic oscillators, consistent with the radial and vertical epicyclic motions, 
\begin{equation}
\label{eqn01}
\delta {\ddot r} + \omega_r^2\,\delta r = 0,~~~
\delta {\ddot z} + \omega_z^2\,\delta z = 0,
\end{equation}
where $\omega_r$ and $\omega_z$ are the radial and vertical epicyclic frequencies. In Newton's gravity with the spherically symmetric potential it is $\omega_r\,$$ =\, $$\omega_z \,$$= \,$$\omega_K$, where $\omega_K$ is the Kepler's orbital frequency, but in strong gravity (see Figure \ref{fig02}),
\begin{equation}
\label{eqn0?}
\omega_z \ge \omega_K > \omega_r, 
\end{equation}
and $\omega_r(r_{\rm MS})\,$$=$$0$, $\omega_r^2\,$$<\,$$0$ for $r\,$$<\,$$r_{\rm MS}$, where $r_{\rm MS}=$ ISCO is the radius of the marginally stable circular orbit. 

For a nearly-Keplerian {\it fluid} (Shakura-Sunyaev disk, slender torus), with the fluid sound speed $C^2_S/\omega_k^2 r^2 \equiv \beta \ll 1$, the frequencies of fluid's epicyclic oscillations ${\bar \omega}_r,~{\bar \omega}_z$ are modified by the pressure\footnote{As first explained in the QPO context by Klu{\'z}niak \& Abramowicz (2002). Discussion that follows is copied from their paper.},
\begin{equation}
\label{eqn02}
{\bar \omega}_r = \omega_r - \chi_r \beta,~~~
{\bar \omega}_z = \omega_z - \chi_z \beta,~~~
\chi_r, \chi_z = {\rm const}
\end{equation}
The modification is very small, but crucial, as it provides a weak pressure coupling between the epicyclic modes of fluid oscillations that leads to the $3/2$ resonance. To see how it works, one writes (\ref{eqn01}) and (\ref{eqn02}) in Mathieu's form, 
$\delta {\ddot z} + \omega_z^2\,[ 1 + \chi \cos ({\bar \omega_r}t)] \delta z = 0$,
$\delta {\ddot r} + {\bar \omega_r}^2\,\delta r = 0$.
A pa\-ra\-met\-ric re\-so\-nan\-ce occurs when ${\bar \omega_r}$$=$$2\,\omega_z/n$, $n$$=$ $1,2,3,...$ In strong gravity ${\bar \omega_r}$$<$$\,\omega_z$, and thus $n$$=$$3$ is the lowest possible value of $n$, corresponding to the strongest resonance. This explains why the ratio, ${\omega_z}/{{\bar \omega_r}} = {\nu_{\rm up}}/{\nu_{\rm lo}} = {3}/{2}$ is most often observed.
\subsection{Mathematical formulation of the model}
The Klu{\'z}niak \& Abramowicz QPO resonance model starts from $\nabla_i\,T^i_{~k} = 0$. Physics, however complicated, is fully described by stress-energy tensor $T^i_{~k}(t) = {}^0T^i_{~k} + {\cal N}^i_{~k}(t)$. The ``mean-field'', statistical average of the flow is described in terms of a slowly varying tensor ${}^0T^i_{~k}$. Turbulent stresses (e.g. the MRI turbulence), fluctuate in time and are described by ${\cal N}^i_{~k}(t)$. A similar division of $T^i_{~k}$ was employed, for a different purpose, by Ogilvie (2003). Deviations from a quasi-Keplerian flow are described by 
$\delta \nabla_i\,T^i_{~k} = 0$, which is put into, 
\begin{eqnarray}
\label{eqn03}
\delta {\ddot r} + \omega_r^2\,\delta r &=& 
F(\delta r, \delta z, \delta {\dot r}, \delta {\dot z}) +
A \cos (\omega_0 t) + 
{\cal N}(t), \nonumber \\
\delta {\ddot z} + \omega_z^2\,\delta z &=& 
G(\delta r, \delta z, \delta {\dot r}, \delta {\dot z}) +
B \cos (\omega_0 t) + 
{\cal N}(t).
\end{eqnarray}
Here the functions $F, G$ account for coupling (known is terms of expansion of $\delta \nabla_i{}^0T^i_{~k}=0$), the $\cos(\omega_0 t)$ terms describe external forcing (relevant in the NS case, $\omega_0$ is the NS spin), and ${\cal N}(t)$ describes an influence of the MRI turbulence. Obviously, $\cos(\omega_0 t)$ and ${\cal N}(t)$ feed energy into the resonance.

Equation (\ref{eqn03}) was studied by the method of multiple scales. All variables,
including time, were expanded according to $X = \sum \epsilon^k X_k$, $\epsilon \ll 1$, and equations of different orders of $\epsilon$ solved separately. Some of the coefficient in expansions have been assumed ad hoc, as it is difficult to derive all of them from first principles. 

The $3/2$ resonance was found to be a natural and robust property of nearly Keplerian fluid motion is strong gravity, and resulting from it double peak QPO frequencies have been found to behave as these observed, in particular the Bursa line was recovered, see Figure \ref{fig01}. 

Obviously, one wants derive all expansion coefficient from first principles. Of a particular interest is the order ${\cal N}(t) \sim \epsilon^N$ that should follow from a quantitative understanding the MRI turbulence. A pure mathematical reasoning hints how a physically interesting range of $N$ may be found.  When $N$ is too large the turbulence is too small and supplies not enough energy to keep the resonance alive. When $N$ is too small, the turbulence is too strong and dominate dynamics, preventing QPO from showing up above the noise. Thus $N$ must be in a particular range, that may be determined e.g. by stochastic differential equations (SDE) that we use. One hopes to compare this range with that predicted by MRI turbulence.
\section{Forced oscillations: QPO and NS spin}            
In in some NS sources, the QPO frequencies obviously depend on the NS spin. For example, difference in frequencies of the double peaked QPO in the millisecond pulsar SAX J1808.4-3658 is clearly equal to half of the pulsar spin. This suggests that the epicyclic resonance is excited by direct forcing of accretion disk oscillation modes by the neutron star spin. 

We studied the forcing in the case of slender tori around rotating NS. All possible linear normal modes of oscillations of a perfect fluid slender torus are known analytically (Blaes, 1985). They include the ``epicyclic'' fluid modes described by equation (\ref{eqn03}), also in the case of a slightly non-slender torus. We study analytically non-linear behavior of these modes, including their coupling and resonances, MRI turbulence, and direct external forcing $\cos(\omega_0 t)$ by the NS rotation $\omega_0$. This is a rather difficult mathematical problem that takes time --- first results are excepted this year\footnote{Our unpublished results, both analytic and numerical, revealed a remarkably complex network of mode coupling and resonances that occur when forcing is included. A similar numerical research of oscillations of fluid tori is being conducted by Omer Blaes and Chris Fragile at UCSB. Rezzolla \& al. (2003) were first to study oscillations of tori in the QPO context.}. So far, all our published results on the $\cos(\omega_0 t)$ external forcing of oscillations of tori are based on numerical simulations. 

It was found that a resonant response occurs when the difference between frequencies of the two epicyclic modes equals to one-half of the spin frequency (as observed in SAX J1808.4-3658 and other "fast rotators"), and when it equals to the spin frequency (as observed in "slow rotators" like XTE J1807-294).

\section{Spectral states, QPO, modulation of X-rays}      
Jean-Pierre Lasota told me recently, reflecting on the ob\-ser\-ved puzzling connections between QPO and spectral states of LMXRB --- ``I am convinced that resonances contribute only a small part to something that is more fundamental there, and connected to how does the disk radiate...\footnote{Jestem przekonany, {\.z}e rezonanse s{\c a} tylko ma{\l}ym przyczynkiem do czego{\'s} bardziej podstawowego, zwi{\c a}zanego z tym jak dysk {\'s}wie\-ci, a MRI (na razie) nie modeluj{\c a} dysk{\'o}w i nie m{\'o}wi{\c a} nic o ich {\'s}wieceniu. Mo{\.z}e za dziesi{\c e}{\'c} lat symulacje b{\c e}d{\c a} przedstawia{\l}y dyski. Takie jest te{\.z} zdanie Boba W.}''. Nobody knows how to explain the puzzle, and we are all guessing. My intuition tells that in both NS and BH sources the QPO clock must be regulated by accretion disk oscillations that are set by strong-gravity's orbital dynamics. In my opinion, how does the matter radiate cannot be fundamentally important for the QPO clock. However, the observational appearance of QPO must be influenced by the same processes that modulate radiation, and this is where I could see a natural link between QPO and spectral states. To be more specific, let me start from explaining how some differences in appearance of QPO in BH and NS sources may follow from the ``unique gravity's clock --- non-unique modulation'' principle. Close to ISCO, the vertical oscillation $\delta z (t)$ modulate ${\dot M}_{in}$ which is very sensitive to the mariginally stable ISCO conditions,
\begin{equation} ({\delta\dot M}/{\dot{M}})_{in}= \chi {\delta z^2},
~~~\chi = \chi (r_{in}, n) = {\rm const}.
\end{equation}
In NS sources, the total X-ray luminosity $L_X$ is dominated by that of the boundary layer, $L_X \sim L_X^{BL} \sim{\dot M}_{in}$. Thus, the QPO modulation of the luminosity occurs at the boundary layer.

For BH this mechanism does not work because the modulated inner mass flux $\dot{M}_{\rm in}$ disappears inside the BH event horizon, and thus no (or only a very little) modulation of the X-ray luminosity may be expected. However, it was shown by a large scale ray tracing method that in this case, the X-ray luminosity is modulated by strong gravity effects (gravitational lensing and Doppler beaming) in photon propagation from oscillating disk to observer. It was found that the ``observed'' power in PDS Fourier spectra very strongly depends on lensing and beaming effects --- for example, power in the vertical mode increases with inclination. 

The ray tracing simulations proved that observed PDS power of QPO modes depends also on the vertical size of the oscillating region of the disk. Obviously, the size varies with temperature that is an important parameter for spectral states. This provides an (unavoidable, based only on photon propagation in strong gravity) link between spectral states and QPO. Note, that it is believed that in different spectral states the location of inner radius of the optically thick Shakura-Sunyaev disk varies. This provides an opaque varying screen that blocks some of the photons involved in lensing and beaming, with important consequences for the QPO visibility in different spectral states.  

Of course, QPO appearance may also be {\it intrinsically} influenced by different physical conditions of the flow in different spectral states. This we study by a direct ray tracing from Mami Machida's accretion flow MHD 3D simulations in
Schwarzschild spacetime. 
\section{Conclusions}                                     
Similarities in QPO observed in BH, NS and WD sources point to a common origin --- accretion disk oscillations. Differences reflect different masses and sizes of BH, NS and WD that induce differences in excitation of oscillations and modulation of luminosity.

In particular, repeatability and precision of the two QPO clocks in double peak QPO in BH and NS sources is explained by a resonance between two epicyclic modes, vertical and radial, that exist in all nearly Keplerian fluid accretion flows. The $3/2$ frequency ratio and the $1/M$ scaling are both a direct and natural (unavoidable) consequence of the strong gravity. The resonance is fed by MRI turbulence in BH disks, and may be additionally forced by rotation of the central star in NS sources. Modulation of X-ray luminosity occurs at the boundary layer in NS sources, and is due to lensing and Doppler beaming in BH sources. Observed correlations between QPO and the LMXRB spectral states do not necessarily imply that QPO strongly depend on radiative processes --- more likely the correlations are determined by the QPO ``visibility'' during different spectral states.

Nature and properties of double peak QPO in BH and NS sources, profoundly depend on Einstein's strong gravity, and on local physics of the MRI turbulence. For this reason, a deeper understanding of QPO may provide a rescue from confusions that upset the accretion theory today. 


\end{document}